\documentstyle[prd,aps,tighten,floats,epsf,epic]{revtex}

\begin{document}
\setlength{\unitlength}{1mm}

\newcommand{\ba} {\begin{eqnarray}}
\newcommand{\ea} {\end{eqnarray}}
\newcommand{\be}{\begin{equation}}
\newcommand{\ee}{\end{equation}}
\newcommand{\n}[1]{\label{#1}}
\newcommand{\eq}[1]{Eq.(\ref{#1})}
\newcommand{\ind}[1]{\mbox{\tiny{#1}}}
\renewcommand\theequation{\thesection.\arabic{equation}}

\newcommand{\la}{\langle}
\newcommand{\ra}{\rangle}

\newcommand{\nn}{\nonumber \\ \nonumber \\}
\newcommand{\nl}{\\  \nonumber \\}
\newcommand{\pr}{\partial}
\renewcommand{\vec}[1]{\mbox{\boldmath$#1$}}

\title{
{\hfill {\small Alberta-Thy-14-00 } } \vspace*{4cm} \\
{\Large \bf Non-Minimally Coupled Massive Scalar Field \\
in a 2D Black Hole: Exactly Solvable Model}}
\bigskip

\author{V. Frolov${}^{(a)}$\thanks{Electronic address: 
frolov@phys.ualberta.ca} 
and 
A. Zelnikov${}^{(a,b)}$\thanks{Electronic address: 
zelnikov@phys.ualberta.ca}
}
\address{$^{(a)}$Theoretical Physics Institute, Department of Physics,
University of Alberta,
Edmonton, AB, Canada T6G 2J1\\
$^{(a)}$P.N.~Lebedev Physics Institute,
Leninsky pr. 53, Moscow  117924, Russia}
\maketitle
\vskip 2cm

\begin{abstract}
We study a nonminimal massive scalar field in a
2-dimensional black hole spacetime. 
We consider the black hole which is the solution of the 2d dilaton 
gravity derived from string-theoretical models.   
We found an explicit solution in a closed form for all modes 
and the Green function
of the scalar field with an arbitrary mass and a nonminimal coupling 
to the curvature. Greybody factors, the Hawking radiation, and
$\la\varphi^2\ra^{\ind{ren}}$ are
calculated explicitly for this exactly solvable model.\\
\bigskip

\noindent
PACS number(s): 03.70.+k, 04.62.+v, 04.70.Dy, 11.10.-z
\end{abstract}


\baselineskip=.6cm

\newpage
\section{Introduction} 
Study of quantum effects in a black hole spacetime is a very
complicated problem. So, an exactly solvable model which exhibits 
the same qualitative properties as the real black hole 
could be extremely useful for understanding quantum physics 
in the real black hole background. First simplification 
frequently used in the literature is the restriction of the metric and
matter fields to have spherical symmetry. The s-wave sector 
contains almost all features of black holes that makes them 
such nontrivial objects. Even quantitatively for the Schwarzschild
black hole about 90\% of Hawking radiation of scalar field is in s-mode.
In spite of an apparent simplicity of
the s-wave sector, the problem of field quantization  
on a generic black hole background still remains quite involved.
There were developed many powerful methods to approach the problem
\cite{MuWiZe,NoOd,BaFa,KuVa,KuVass-99,FroSutZel,BFFNSZ-2000}. 
It would be nice to have a good test for their accuracy and
limitations of applicability. 

The 2d gravity coupled to a dilaton $\phi$ field with action
\begin{equation}
S_g={1\over 2\pi}\int d^2 x\sqrt{g}{\mathrm e}^{-2\phi}\left[
R+4(\nabla\phi)^2+4\lambda^2\right] \label{S_g}
\end{equation}
has black hole solutions with the desired properties.
First of all, the classical solutions are known in explicit closed form.
This action arises in a low-energy asymptotic of string theory models
\cite{ManSenWad,Witten,CGHS,Strom-95,RuTseyt,Frolov-92,McGuiNappYost} 
and in
certain  models with a scalar matter \cite{Mann,CanJackiw,Chams}
(For the review of more general dilaton gravity models see 
\cite{Strobl} and references therein)
The corresponding black hole solution has similar properties to those
of the $(r,t)$ sector of the Schwarzschild black hole and the 
Carter-Penrose conformal diagram is also similar. 
Quantum fields propagating on this background are very similar
to the Schwarzschild case but appear to be much simpler to deal with.
Most of the papers on this subject consider conformal matter
on the 2d black hole spacetimes. 

In this paper we address to the problem of quantization of
non-conformal fields. 
This problem is more complicated but much more
interesting since nonconformal fields interact with the curvature
and feel the potential barrier, which also plays an important role 
in black hole physics. In this case it is important to know greybody 
factors to study, e.g., the Hawking radiation and vacuum polarization 
effects. For the minimally coupled scalar fields in 2d there is no 
potential barrier and greybody factors are trivial.
Qualitative features of the potential barrier of the Schwarzschild
black hole are very close to that of the string inspired 
2d gravity model we consider here.

\section{Model}
\setcounter{equation}0

Our purpose is to study a quantum scalar massive field in a spacetime of
the 2-dimensional black hole. The most general static 2-dimensional
metric can be written in the form
\be \n{2.1}
dS^2 = -f dT^2 +{dr^2\over f}\, ,
\ee
For the theory (\ref{S_g}) the function $f$ and the corresponding
solution for the dilaton field $\phi$ are
\be\n{2.2}
f=1-e^{-{r\over r_0}}\, , \hskip 1cm \phi={r\over r_0}\,.
\ee
This solution of the dilaton gravity action (\ref{S_g}) describes
the string-theoretic black hole \cite{ManSenWad,Witten} . 
Like the Schwarzschild black hole it is asymptotically flat at $r=\infty$.
The metric contains only one parameter $r_0$, which determines the position of
the horizon. The surface gravity $\kappa$ and the scalar curvature $R$ are 
\be\n{2.3}
\kappa ={1\over 2} {df\over dr}|_{r=r_0}={1\over 2r_0}\, ,\hspace{0.5cm}
R=-{d^2 f\over dr^2}={1\over r_0^2} e^{-r/r_0}\, .
\ee

It would be convenient to use the dimensionless form of (\ref{2.1})
\be\n{2.4}
ds^2=r_0^{-2}\, dS^2 = -f dt^2 +{dx^2\over f}\, ,
\ee
where $t=T/r_0$, $x=r/r_0$, and
\be\n{2.5}
f=1-e^{-x}\, .
\ee
The dimensionless surface gravity, $\tilde{\kappa}$, and curvature,
$\tilde{R}$, are
\be\n{2.6}
\tilde{\kappa}={1\over 2}\, ,\hspace{1cm}\tilde{R}=e^{-x}\, .
\ee
By introducing a new variable
\be\n{2.6a}
z=1-e^{-x}\, ,
\ee
the metric can be written in an algebraic form
\be\n{2.6b}
ds^2= -z\, dt^2 +{dz^2\over z\, (1-z)^2}\, .
\ee
We shall also use another form of the metric
\be\n{2.6c}
ds^2 =-z\, dw_{\pm}^2 \pm \, {2\,dw_{\pm}\, dz\over 1-z}\, ,
\ee
where $w_{\pm}=t \pm \ln(z/(1-z))$.  In these coordinates
the curvature $\tilde{R}=1-z$.

The field equation follows from the action
\be\n{2.7}
W[\varphi ]=-\frac{1}{2}\int
dX^2\,g^{1/2}\,\left[\left(\nabla\varphi\right)^2
+(m^2+\xi R)\,\varphi^2\right]\, .  
\ee
and has the form
\be\n{2.8}
D_{m,\xi}\varphi =0\, ,\hspace{0.5cm}D_{m,\xi}=\Box -m^2 -\xi R\, .
\ee
In these relations $\xi$ is a parameter of non-minimal coupling. The 
stress-energy tensor, $T_{\mu\nu}$, for the field reads
\ba
T_{\mu\nu}&\equiv&{2\over \sqrt{g}}{\delta W\over \delta g^{\mu\nu}}
= (1-2\xi )\,\varphi_{\!,\mu}\,\varphi_{\!,\nu} 
\nonumber \\
&&+ \left(2\xi-\frac{1}{2}\right)
g_{\mu\nu}\,g^{\alpha\beta}\,\varphi_{\!,\alpha}\,\varphi_{\!,\beta }
\nonumber \\
&&-2\xi \varphi \left( \varphi_{;\mu\nu} - 
g_{\mu\nu}\,\Box \varphi\right) \nonumber \\
&&+\xi \left(R_{\mu\nu} - \frac{1}{2}g_{\mu\nu}\,R\right)
\varphi^2 -{1\over 2}m^2\varphi^2 g_{\mu\nu}.  \n{2.9}
\ea
The field equation (\ref{2.8}) can be written as
\be\n{2.10}
\left[-{\partial^2\over \partial t^2} + {\partial^2 \over \partial
x_*^2} -U \right] \varphi =0 \, ,
\ee
where $\mu=mr_0$,
\ba
U&=&f(x)\, ({\mu}^2 +\xi R) \nonumber \\
 &=& {{\mu}^2+\xi\over 1+\exp(-x_*)}-{\xi\over 
(1+\exp(-x_*))^2}\, ,
\n{2.11}
\ea
and $x_*$ is ''tortoise'' coordinate
\be\n{2.12}
x_{*}\equiv \int_1^x{dx\over f(x)}=\ln(e^x-1)\, .
\ee

\begin{figure}
\epsfysize=8 cm
\centerline{\epsfbox{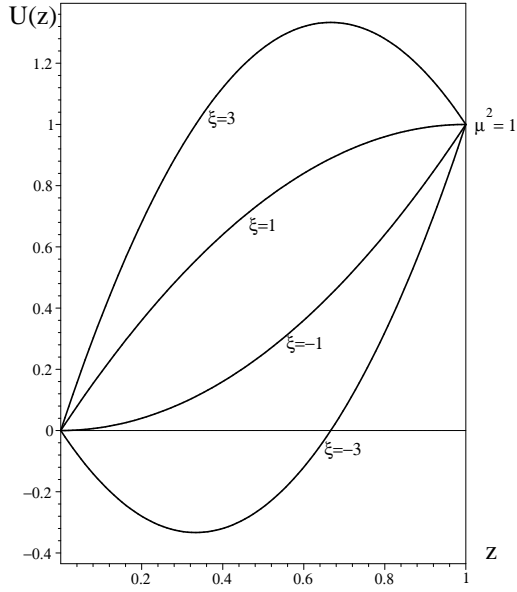}}
\caption[fig1]
{Potential $U(z)$ for $\mu^2=1$ as a function of $z$. 
Horizon is at $z=0$.}
\label{fig1}
\end{figure}

Solutions of this equation can be constructed from monochromatic waves
$\varphi \sim {\mathrm e}^{-i\omega t} R(x|\omega)$ where ``radial'' modes
$R(x|\omega)$ obey the equation 
\be\n{2.13} 
\left[{\partial^2 \over
\partial x_*^2} +\omega^2-U \right]  R(x|\omega)=0 \, . 
\ee 
Properties
of solutions depend on the form of the potential $U$. Using the
relation 
\be\n{2.14} 
z=1-\exp(-x)=1/(1+\exp(-x_*)\, , 
\ee 
one gets
\be\n{2.15} 
U(z)=z( (\mu^2+\xi) -\xi z)\, . 
\ee 
Thus 
\be\n{2.16}
U'(z)=(\mu^2+\xi) -2\xi z\, ,\hspace{0.5cm} U''(z)=-2\xi\, . 
\ee
Function $U(z)$ has maximum (minimum) inside the interval $z\in (0,1)$
when $\xi >\mu^2$ ($\xi < -\mu^2$). For $|\xi| < \mu^2$ function $U(z)$
is a monotonic function in the same interval. Qualitatively its
behavior is presented in Figure \ref{fig1}. 

Let us consider the case $\xi < -\mu^2$. 
Denote $\eta =-\xi$. The potential 
$U(z)$ is negative in the domain $z\in (0, z_1)$ where 
$z_1=1-\mu^2/\eta$.  For a fixed value of mass $\mu$ 
the depth of the potential well depends on $\xi$. 
The larger $|\xi|$ for negative $\xi$ 
the deeper the well is. Hence there always exist negative 
We consider these states and condition of their formation in Section~4.

\section{Black Hole Radiation}
\setcounter{equation}0

\subsection{Scattering modes}

Let us discuss now the scattering problem for this potential.  As usual,
we introduce the sets of solutions of equation (\ref{2.13}) 
$R_{\rm{up}}$ and $R_{\rm{in}}$
which are specified by their asymptotic behavior as follows
\be\n{2.20}
R_{\rm in}(x_\ast |\omega) \sim {1\over \sqrt{4\pi}} \left\{ \begin{array}{ll}
\omega^{-1/2}\, T_{\omega}\, e^{ -i\omega  x_\ast}\ , 
& x_\ast\rightarrow  -\infty \ , \\ \\
\varpi^{-1/2}\left[R_\omega\,e^{ i\varpi  x_\ast}+ 
\,e^{ -i\varpi  x_\ast}\right]\ , &x_\ast \rightarrow   
+\infty \ ,
\end{array}
\right.
\label{inmode}\ee
\be\n{2.21}
R_{\rm up}(x_*|\omega) \sim {1\over \sqrt{4\pi}} \left\{ \begin{array}{ll}
\omega^{-1/2}\, \left[\,e^{ i\omega  x_\ast}+r_\omega\,e^{ -i\omega 
x_\ast}\right]\ ,
&x_\ast \rightarrow   
-\infty \ ,  \\ \\
\varpi^{-1/2}\, t_{\omega}\, e^{ +i\varpi  x_\ast}\ , & x_\ast\rightarrow  +\infty \ ,
\end{array}
\right.
\ee
where
\be\n{2.22}
\varpi =\sqrt{\omega^2-\mu^2}\, .
\ee
We also denote 
\be\n{2.23}
R_{\rm{out}}(x|\omega)=(R_{\rm{in}}(x|\omega))^*\, ,\hspace{0.5cm}
R_{\rm{down}}(x|\omega)=(R_{\rm{up}}(x|\omega))^*\, .
\ee

\begin{figure}
\centerline{\setlength{\unitlength}{0.00500000in}
\renewcommand{\dashlinestretch}{30}
\begin{picture}(566,608)(0,-10)
\drawline(53,520)(53,520)
\drawline(123,350)(243,470)(123,590)
	(3,470)(123,350)
\dashline{4.000}(203,510)(123,430)
\drawline(198.757,502.929)(203.000,510.000)(195.929,505.757)
\dashline{4.000}(193,520)(123,450)
\drawline(188.757,512.929)(193.000,520.000)(185.929,515.757)
\dashline{4.000}(173,540)(123,490)
\drawline(168.757,532.929)(173.000,540.000)(165.929,535.757)
\dashline{4.000}(183,530)(123,470)
\drawline(178.757,522.929)(183.000,530.000)(175.929,525.757)
\dashline{4.000}(163,550)(123,510)
\drawline(158.757,542.929)(163.000,550.000)(155.929,545.757)
\dashline{4.000}(83,550)(123,510)
\drawline(90.071,545.757)(83.000,550.000)(87.243,542.929)
\dashline{4.000}(73,540)(123,490)
\drawline(80.071,535.757)(73.000,540.000)(77.243,532.929)
\dashline{4.000}(63,530)(123,470)
\drawline(70.071,525.757)(63.000,530.000)(67.243,522.929)
\dashline{4.000}(53,520)(123,450)
\drawline(60.071,515.757)(53.000,520.000)(57.243,512.929)
\dashline{4.000}(43,510)(123,430)
\drawline(50.071,505.757)(43.000,510.000)(47.243,502.929)
\drawline(123,510)(203,430)
\drawline(123,490)(193,420)
\drawline(123,470)(183,410)
\drawline(123,450)(173,400)
\drawline(123,430)(163,390)
\drawline(443,350)(563,470)(443,590)
	(323,470)(443,350)
\drawline(443,30)(563,150)(443,270)
	(323,150)(443,30)
\drawline(123,30)(243,150)(123,270)
	(3,150)(123,30)
\drawline(513,520)(513,520)
\dashline{4.000}(363,510)(443,430)
\drawline(370.071,505.757)(363.000,510.000)(367.243,502.929)
\dashline{4.000}(373,520)(443,450)
\drawline(380.071,515.757)(373.000,520.000)(377.243,512.929)
\dashline{4.000}(393,540)(443,490)
\drawline(400.071,535.757)(393.000,540.000)(397.243,532.929)
\dashline{4.000}(383,530)(443,470)
\drawline(390.071,525.757)(383.000,530.000)(387.243,522.929)
\dashline{4.000}(403,550)(443,510)
\drawline(410.071,545.757)(403.000,550.000)(407.243,542.929)
\dashline{4.000}(483,550)(443,510)
\drawline(478.757,542.929)(483.000,550.000)(475.929,545.757)
\dashline{4.000}(493,540)(443,490)
\drawline(488.757,532.929)(493.000,540.000)(485.929,535.757)
\dashline{4.000}(503,530)(443,470)
\drawline(498.757,522.929)(503.000,530.000)(495.929,525.757)
\dashline{4.000}(513,520)(443,450)
\drawline(508.757,512.929)(513.000,520.000)(505.929,515.757)
\dashline{4.000}(523,510)(443,430)
\drawline(518.757,502.929)(523.000,510.000)(515.929,505.757)
\drawline(443,510)(363,430)
\drawline(443,490)(373,420)
\drawline(443,470)(383,410)
\drawline(443,450)(393,400)
\drawline(443,430)(403,390)
\drawline(53,100)(53,100)
\dashline{4.000}(83,70)(123,110)
\dashline{4.000}(73,80)(123,130)
\dashline{4.000}(63,90)(123,150)
\dashline{4.000}(53,100)(123,170)
\dashline{4.000}(43,110)(123,190)
\drawline(123,150)(183,210)
\drawline(178.757,202.929)(183.000,210.000)(175.929,205.757)
\drawline(123,170)(173,220)
\drawline(168.757,212.929)(173.000,220.000)(165.929,215.757)
\drawline(123,190)(163,230)
\drawline(158.757,222.929)(163.000,230.000)(155.929,225.757)
\drawline(123,110)(203,190)
\drawline(198.757,182.929)(203.000,190.000)(195.929,185.757)
\drawline(123,130)(193,200)
\drawline(188.757,192.929)(193.000,200.000)(185.929,195.757)
\dashline{4.000}(163,70)(123,110)
\drawline(130.071,105.757)(123.000,110.000)(127.243,102.929)
\dashline{4.000}(173,80)(123,130)
\drawline(130.071,125.757)(123.000,130.000)(127.243,122.929)
\dashline{4.000}(183,90)(123,150)
\drawline(130.071,145.757)(123.000,150.000)(127.243,142.929)
\dashline{4.000}(193,100)(123,170)
\drawline(130.071,165.757)(123.000,170.000)(127.243,162.929)
\dashline{4.000}(203,110)(123,190)
\drawline(130.071,185.757)(123.000,190.000)(127.243,182.929)
\drawline(513,100)(513,100)
\dashline{4.000}(393,80)(443,130)
\dashline{4.000}(483,70)(443,110)
\dashline{4.000}(493,80)(443,130)
\dashline{4.000}(503,90)(443,150)
\dashline{4.000}(513,100)(443,170)
\dashline{4.000}(523,110)(443,190)
\drawline(443,110)(363,190)
\drawline(370.071,185.757)(363.000,190.000)(367.243,182.929)
\drawline(443,130)(373,200)
\drawline(380.071,195.757)(373.000,200.000)(377.243,192.929)
\drawline(443,150)(383,210)
\drawline(390.071,205.757)(383.000,210.000)(387.243,202.929)
\drawline(443,170)(393,220)
\drawline(400.071,215.757)(393.000,220.000)(397.243,212.929)
\drawline(443,190)(403,230)
\drawline(410.071,225.757)(403.000,230.000)(407.243,222.929)
\dashline{4.000}(363,110)(443,190)
\drawline(438.757,182.929)(443.000,190.000)(435.929,185.757)
\dashline{4.000}(373,100)(443,170)
\drawline(438.757,162.929)(443.000,170.000)(435.929,165.757)
\dashline{4.000}(383,90)(443,150)
\drawline(438.757,142.929)(443.000,150.000)(435.929,145.757)
\dashline{4.000}(403,70)(443,110)
\drawline(438.757,102.929)(443.000,110.000)(435.929,105.757)

\put(78,325){\makebox(0,0)[lb]{\smash{in-mode}}}
\put(398,325){\makebox(0,0)[lb]{\smash{up-mode}}}
\put(73,5){\makebox(0,0)[lb]{\smash{out-mode}}}
\put(383,5){\makebox(0,0)[lb]{\smash{down-mode}}}
\put(23,380){\makebox(0,0)[lb]{\smash{$H^{-}$}}}
\put(343,540){\makebox(0,0)[lb]{\smash{$H^{+}$}}}
\put(343,380){\makebox(0,0)[lb]{\smash{$H^{-}$}}}
\put(23,220){\makebox(0,0)[lb]{\smash{$H^{+}$}}}
\put(343,220){\makebox(0,0)[lb]{\smash{$H^{+}$}}}
\put(23,60){\makebox(0,0)[lb]{\smash{$H^{-}$}}}
\put(343,60){\makebox(0,0)[lb]{\smash{$H^{-}$}}}
\put(23,540){\makebox(0,0)[lb]{\smash{$H^{+}$}}}
\put(193,540){\makebox(0,0)[lb]{\smash{${\cal J}^{+}$}}}
\put(193,380){\makebox(0,0)[lb]{\smash{${\cal J}^{-}$}}}
\put(513,380){\makebox(0,0)[lb]{\smash{${\cal J}^{-}$}}}
\put(513,540){\makebox(0,0)[lb]{\smash{${\cal J}^{+}$}}}
\put(513,220){\makebox(0,0)[lb]{\smash{${\cal J}^{+}$}}}
\put(513,60){\makebox(0,0)[lb]{\smash{${\cal J}^{-}$}}}
\put(193,60){\makebox(0,0)[lb]{\smash{${\cal J}^{-}$}}}
\put(193,220){\makebox(0,0)[lb]{\smash{${\cal J}^{+}$}}}
\end{picture}}
\caption[fig.1]{ {\em IN, UP, OUT} and {\em DOWN}-modes .}
\label{fig.1}
\end{figure}
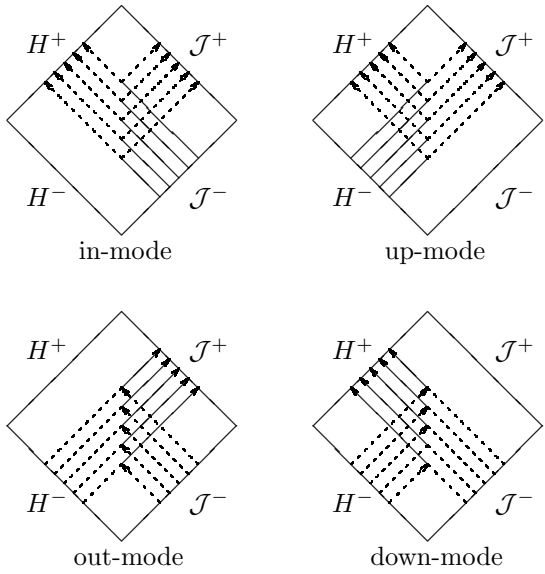

Modes $R_{\rm in}$ and $R_{\rm out}$ are defined for 
$\omega >\mu$, while modes $R_{\rm up}$ and $R_{\rm down}$ are
defined for $\omega >0$. It should be emphasized that for $0<\omega<\mu$
we define $\varpi=i\sqrt{\mu^2 -\omega^2}$, so that in this domain of
frequencies modes $R_{\rm up}$ are exponentially decreasing at large
values of $x_*$. In the absence of bound states the modes 
$R_{\rm in}$ and $R_{\rm up}$ and their
complex conjugated form a complete set (basis). We consider this case
first and discuss aspects of the problem connected with the existence of
bound states later.

Evaluating Wronskians for these solutions at $x_*=\pm \infty$, one can
show that for $\omega >\mu$ 
\be\n{2.24}
|R_{\omega}|^2+|T_{\omega}|^2= 1\, ,
\ee
\be\n{2.25}
|r_{\omega}|^2+|t_{\omega}|^2= 1\, ,
\ee
and
\be\n{2.26}
T_{\omega}=t_{\omega}\, .
\ee
For $0<\omega < \mu$ a wave emitted from the horizon does not reach an
infinity and is totally reflected. For these waves $|R_{\omega}|=1$. 

Only two of four solutions, $R_{\rm{up}}$, $R_{\rm{in}}$,
$R_{\rm{down}}$, and $R_{\rm{out}}$, are linearly independent. By using
asymptotics of these solution one can easily obtain the following
relations
\ba\n{2.27}
R_{\rm{in}}(x|\omega)&=&T_{\omega}\,  R_{\rm{down}}(x|\omega)+
 R_{\omega}\, R_{\rm{out}}(x|\omega)\, .
\\
\n{2.28}
R_{\rm{up}}(x|\omega)&=&r_{\omega}\,  R_{\rm{down}}(x|\omega)+
t_{\omega}\, R_{\rm{out}}(x|\omega)\, ,
\ea
Other similar relations can be obtained by applying complex conjugation
to (\ref{2.27}) and (\ref{2.28}).

We demonstrate now that a general solution of the field equation
(\ref{2.13}) can be obtained in terms of hypergeometric functions. For
this purpose we write $R(x|\omega)$ in the form
\be\n{2.29}
R(x|\omega)=q_{\omega}(z)\, u(z|\omega)\, ,
\ee
where
\be\n{2.30}
q_{\omega}(z)= z^{-i\omega}\, (1-z)^{-i\varpi}\, .
\ee
One can check that the function $u(z|\omega)$ obeys the following
hypergeometric equation
\be\n{2.31}
z(1-z)\, {d^2 u\over dz^2}+[c-(a+b+1)z]\, {d u\over dz} -ab\, u=0\, ,
\ee
where
\be\n{2.32}
a=\alpha -\beta\, ,\hspace{0.5cm}
b=\alpha+\beta\, ,\hspace{0.5cm}c=1-2i\omega\, ,
\ee
\be\n{2.33}
\alpha={1\over 2}-i(\omega+\varpi)\, ,\hspace{0.5cm}
\beta=\sqrt{{1\over 4}-\xi}\, .
\ee
Consider first the following two linearly independent solutions 
\be\n{2.34}
u_1=F(a,b;c;z)\, ,\hspace{0.5cm}
u_2=F(a,b;\tilde{c};1-z)\, ,
\ee
where
\be\n{2.35}
\tilde{c}=1-2i\varpi\, .
\ee
Here and later we use the same notations $u_i$ for special solutions of
the hypergeometric equation as in the section 2.9 of the book
\cite{BeEr}.  Solutions $R_{\rm up}$ and $R_{\rm in}$ which have have  
asymptotics (\ref{2.20}) and (\ref{2.21}) can be written as
\ba\n{2.36}
R_{\rm in}(x|\omega)&=&{T_{\omega}\over \sqrt{4\pi\omega}}\,
q_{\omega}u_1\equiv
{T_{\omega}\over \sqrt{4\pi\omega}}\,
q_{\omega}(z)\,F(a,b;c;z)\, , 
\\
\n{2.37}
R_{\rm up}(x|\omega)&=& 
{t_{\omega}\over \sqrt{4\pi\varpi}}\,
q_{\omega} u_2\equiv
{t_{\omega}\over \sqrt{4\pi\varpi}}\,
q_{\omega}(z)\,F(a,b;\tilde{c};1-z)\, . 
\ea
We use also two other different solutions of the hypergeometric equation
denoted by $u_5$ and $u_6$ in \cite{BeEr} (section 2.9)
\ba\n{2.38}
u_5&=&z^{1-c}\, (1-z)^{c-a-b}\, F(1-a,1-b;2-c;z)\, ,
\\
\n{2.39}
u_6&=& z^{1-c}\, (1-z)^{c-a-b}\, F(1-a,1-b;c+1-a-b;1-z)\, .
\ea
Notice now that
\be\n{2.40}
a^* =  \left\{ \begin{array}{ll}
1-b\ , 
& \mbox{for real } \beta  \ , \\ \\
1-a , & \mbox{for imaginary } \beta\ ,
\end{array}
\, ,\hspace{0.5cm}
b^* =  \left\{ \begin{array}{ll}
1-a\ , 
& \mbox{for real } \beta  \ , \\ \\
1-b , & \mbox{for imaginary } \beta\ .
\end{array}
\right.\, \right. 
\ee
Using these relations and symmetry of the hypergeometric function with
respect to its first two arguments we get
\ba\n{2.41}
u_5&=&z^{2i\omega}\, (1-z)^{2i\varpi}\, F(a^*,b^*;c^*;z)\, ,
\\
\n{2.42}
u_6&=&z^{2i\omega}\, (1-z)^{2i\varpi}\, F(a^*,b^*;\tilde{c}^*;1-z)\, .
\ea
Thus we have
\be\n{2.43}
R_{\rm out}= {T^*_{\omega}\over \sqrt{4\pi\omega}}\,\,
q_{\omega}\, u_5\, ,\hspace{0.5cm} 
R_{\rm down}= 
{t^*_{\omega}\over \sqrt{4\pi\varpi}}\,\,
q_{\omega}\,  u_6\, .
\ee

\subsection{Transition and reflection coefficients}

To determine transition and reflection coefficients it is sufficient to
use Kummer relations between $u_i$. In particular one has (see equation
(2.9.41) of \cite{BeEr})
\be\n{2.44}
u_5= A\, u_2 +B\, u_6\, ,
\ee
where
\be\n{2.45}
A={\Gamma(2-c)\, \Gamma(c-a-b)\over \Gamma(1-a)\, \Gamma(1-b)}\,
,\hspace{0.5cm}
B={\Gamma(2-c)\, \Gamma(a+b-c)\over \Gamma(a+1-c)\, \Gamma(b+1-c)}\, .
\ee
By comparing relation (\ref{2.44}) with (\ref{2.28}) one gets
\be\n{2.46}
T_{\omega}=\sqrt{ {\omega\over \varpi}}\, {1\over A^*}\, ,\hspace{0.5cm}
r_{\omega}=-{t_{\omega}\over t^*_{\omega}}\, {B\over A}\, .
\ee
Coefficients $R_{\omega}$ and $t_{\omega}$ can be obtained in a similar
manner.

After simple transformations the coefficient $T_{\omega}$ can be
presented as
\be\n{2.47}
T_{\omega}=\sqrt{ {\omega\over \varpi}}\,{\Gamma(a)\,
\Gamma(b)\over \Gamma(c)\, \Gamma(\tilde{c}-1)}\, . 
\ee
By using relations
\be\n{2.48}
\Gamma(\zeta)\, \Gamma(1-\zeta)={\pi\over \sin(\pi\zeta)}\,
,\hspace{0.5cm}
\Gamma(\zeta)\, \Gamma(-\zeta)=-{\pi\over \zeta\sin(\pi\zeta)}\, ,
\ee
we can write $|T_{\omega}|^2$ in the form
\be\n{2.49}
|T_{\omega}|^2 ={2\sinh(2\pi\omega)\, \sinh(2\pi\varpi)\over
\cosh[2\pi(\omega +\varpi)] +\cos(2\pi\beta)}\, .
\ee
This relation is valid both for real and imaginary $\beta$. It is periodic
in $\beta$ with a period 1.

\subsection{Radiation spectrum and energy flux}

\begin{figure}
\epsfysize=8 cm
\centerline{\epsfbox{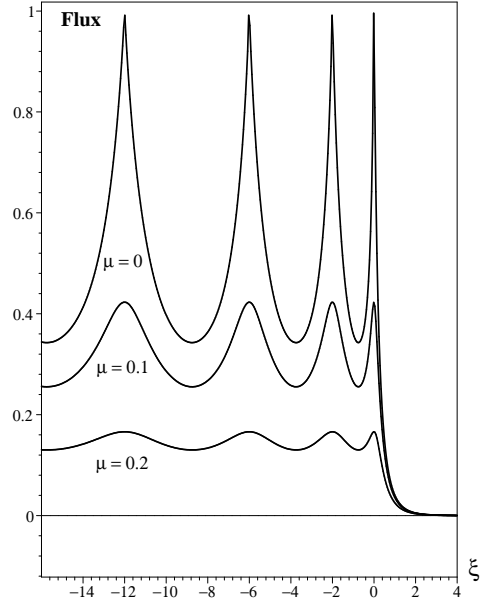}}
\caption[fig3]
{The  
$\mbox{\rm\bf Flux}=192\pi r_0^2\times{dE\over dt}$  
as a function of $\xi$ (for $\mu=0,~\mu=0.1$,
and $\mu=0.2$).}
\label{fig3}
\end{figure}

The number density of particles radiated by the black hole to infinity
in the range of frequencies $(\omega,\omega+d\omega)$ is given by Hawking
expression
\be\n{2.50}
{dn(\omega)\over d\omega}={|T_{\omega}|^2\over \exp(4\pi\omega)-1}=
{\exp(-2\pi\omega)\sinh(2\pi\varpi)\over
\cosh[2\pi(\omega +\varpi)] +\cos(2\pi\beta)}\, .
\ee
The corresponding energy density flux is (see Figure \ref{fig3})
\be\n{2.51}
{dE\over dt}={1\over 2\pi r^2_0}\, \int_{\mu}^{\infty}{d\omega\, \omega\,
\exp(-2\pi\omega)\sinh(2\pi\varpi)\over
\cosh[2\pi(\omega +\varpi)] +\cos(2\pi\beta)}\, .
\ee
(We restored the dimensionality in this relation.)
For $\mu=0$ this integral can be calculated exactly 
\ba
\left. {dE\over dt}\right|_{\mu=0}&=&{1\over 32\pi^3 r^2_0}\,{e^{2\pi i \beta}\,
\mbox{dilog}(1+e^{-2\pi i\beta})-\mbox{dilog}(1+e^{2\pi i\beta}) \over
e^{2\pi i\beta}-1}\, \nonumber\\
&\equiv&{1\over 32\pi^3 r^2_0}\,{e^{\pi\sqrt{4\xi-1}}\,
\mbox{dilog}(1+e^{-\pi\sqrt{4\xi-1}})-\mbox{dilog}(1+e^{\pi\sqrt{4\xi-1}}) \over
e^{\pi\sqrt{4\xi-1}}-1}\,
\n{2.52}.
\ea
Here $\mbox{dilog}(z)$ is the dilogarithm function
\be\n{2.53}
\mbox{dilog}(z)=-\int_1^z \, {\ln t\over t-1}\, dt\, .
\ee
Let us emphasize that in the above calculations of the energy flux we
excluded a contribution of possible bound states, which are discussed
in the next section.

\section{Bound State and Black Hole Instability}
\setcounter{equation}0

\subsection{Bound states}

Solving the field equation we assumed that $\omega$ is real. Besides
these wave-like solutions the system can have modes with time dependence
$\sim \exp(\pm \Omega t)$. For this modes
\be\n{4.1}
q_{\Omega}(z)=z^{\Omega}\, (1-z)^{\tilde{\Omega}}\, ,
\ee
\be\n{4.2}
\tilde{\Omega}=\sqrt{\Omega^2+\mu^2}\, ,
\ee
and a corresponding ``radial'' function is
\be \n{4.3}
R(x|\Omega)=q_{\Omega}(z)\, u(z|\Omega)\, .
\ee
If $u(z|\Omega)$ is finite both at $z=0$ and $z=1$, the radial function
for $\Re(\Omega)>0$  has decreasing asymptotics both at the horizon and
infinity. We shall call these states bound states. 

As usual we denote by $u_1$ a solution which is finite at $z=0$ and by
$u_2$ a solution which is finite at $z=1$. These solutions are given by
relation (\ref{2.34}) with
\be\n{4.4}
a= {1\over 2}+\Omega +\tilde{\Omega}-\sqrt{{1\over 4}-\xi}\, ,
\hspace{0.5cm}
b= {1\over 2}+\Omega +\tilde{\Omega}+\sqrt{{1\over 4}-\xi}\, ,
\ee
\be\n{4.5}
c=1+2\Omega \, ,\hspace{0.5cm}\tilde{c}=1+2\tilde{\Omega}\, .
\ee
A bound state exists if for
a given $\Omega$ solutions $u_1$ and $u_2$ are linearly dependent. It
follows from
Kummer relations (see (2.9.25) and (2.9.26) of \cite{BeEr}) that the
solutions are linearly dependent if and only if either $a$ or $b$ is a
non-positive integer number. For $\Re(\Omega)>0$ this condition can be
satisfied only if both $\beta\equiv\sqrt{1/4-\xi}$ and $\Omega$ are real. A
condition which determines $\Omega$ is
\be\n{4.6}
\Omega_n +\tilde{\Omega}_n= B-n\, ,\hspace{0.5cm}
B\equiv\beta-{1\over 2}\, ,
\ee
where $n\ge 0$. Bound states are possible only if the non-minimal
coupling constant $\xi$ is negative. The number of bound states is
defined by the integer part of the quantity $[B-\mu+1]$. 
Solving (\ref{4.6}) we get
\be\n{4.7}
\Omega_n ={1\over 2}\, \left[ (B-n)-{\mu^2\over B-n}\right]\, .
\ee

When the parameter $\xi$ is negative and decreases further, the
number of bound states increases. A new bound state appears when
$B-\mu$ reaches a new integer number value. In particular, transition
from a pure continuous spectrum to a spectrum with a 
single bound state occurs when $B=\mu$ or
\be\n{4.8}
-\xi =\mu (\mu+1)\, .
\ee

The condition $a=-n$ which determines a bound states implies that the
series in the definition of the hypergeometric function is truncated and
in fact the corresponding solution $u=f_n$ is a polynomial of power $n$
\be\n{4.9}
f_n(z)=\sum_{k=0}^n {(-n)_k\, (b_n)_k\, z^k\over (c_n)_k\, k!}\, ,
\ee
where $(p)_0=1$ and $(p)_k= p(p+1)...(p+k-1)$ for $k=1,2,3,\ldots$ \ .
In particular, we have
\be\n{4.10}
f_0=1\, ,\hspace{0.5cm}
f_1=1-{b_n\over c_n}\, z\, ,\hspace{0.5cm}
f_2=1-2\, {b_n\over c_n}\, z\, + {b_n\, (b_n+1)\over c_n\, (c_n+1)}\,
z^2\, .
\ee
Since for $\Omega_n\ge 0$ both $b_n\ge 1$ and $c_n\ge 1$, the functions
$f_n$ and their first $n$ derivatives do not vanish at $z=0$. 

\subsection{Energy density and energy fluxes for bound states}

Let us calculate now the stress-energy tensor for a bound state mode.
For our purposes it is convenient to make calculations in the advanced
time coordinates $(v\equiv w_+,z)$, (\ref{2.6c}). For simplicity we
consider the first bound state $n=0$. For this state
\be\n{4.11}
\varphi(v,z)=e^{\Omega_0 v}\, (1-z)^{B}\, .
\ee
We use here the property $\Omega_0+\tilde{\Omega}_0=B$. Calculations
made by using GRTensor give
\ba\n{4.12}
T_{\mu\nu}&=&\varphi^2(v,z) \, t_{\mu\nu}\, , \\
t_{vv}&=&{1\over 4 B^2}\left[
B^4\left((6B+3+4B^2)+(-4+10B+16B^2)z+2(4B+3)(2B+1)z^2\right)
\right. \nonumber \\
&&\left.
-2B^2 \left((5B+2+4B^2)+(B+1)(8B-1)z\right)\,\mu^2
+(2B+1)^2\, \mu^4
\right]\, , \n{4.13} \\
t_{vz}&=&{1\over 2(1-z)}\left[B^2\,
\left((4B^2-2+2B)+(10B+8B^2+3)\,z\right)
+(1-4B-4B^2)\, \mu^2
\right]
\, , \n{4.14} \\
\n{4.15}
t_{zz}&=&{B^2(2B+1)^2\over(z-1)^2}\, .
\ea
We used (\ref{4.7}) to express $\Omega_0$ in terms of $B$ and $\mu$.

The obtained result allows one to show that the energy density flux through the
event horizon, $T_{vv}$ is
\be\n{4.16}
\left. T_{vv}\right|_H={B^2-\mu^2\over 4B^2}\, \exp((B^2-\mu^2)v/B)
\left[ B^2\, (4B^2+6B+3)-\mu^2\, (2B+1)^2\right]\, .
\ee
In the presence of a bound state (when $B>\mu$) 
the energy density flux $\left. T_{vv}\right|_H$ is positive and 
grows exponentially with the advanced time parameter $v$. 
This behavior reflects instability of the quantum system in the
presence of bound states. Quantization of scalar fields  
in the presence of imaginary frequency
modes was discussed by Kang \cite{Kang} (see also references therein).

\section{Green Function}\label{s2}
\setcounter{equation}0

\subsection{Massive field case}

By making Wick's rotation $t\to i\tau$ in the metric (\ref{2.4}) one gets the Euclidean metric of the form
\be
ds_E^2 =f\, d\tau^2 +{dx^2\over f}\, .
\ee
The condition of regularity at $x=1$ requires that the coordinate
$\tau$ has the period $4\pi$.

The Euclidean Green function $G(X,X')$ which contains a complete
information about the quantum field is a solution of the equation
\be\n{n2.10}
\left[ \Box_E -\mu^2-\xi\, \tilde{R}\right]\, G(X,X')=-\delta(X,X')\, .
\ee
The Green function is assumed to be regular at the Euclidean horizon
$r=r_0$ and to vanish at infinity, $r\to\infty$.
Equation (\ref{n2.10}) in the metric (\ref{2.1}) has the following form
\be\n{n2.11}
\left[f^{-1}{\partial^2\over \partial\tau^2} + {\partial \over \partial x}\left(f
{\partial \over \partial x}\right) -{\mu}^2 -\xi \tilde{R}\right]
G(X,X')=-\delta(\tau-\tau')\, \delta(x-x') \, ,
\ee
where $\mu=mr_0$.
This equation allows a separation of variables, so that we can write
\be\n{n2.12}
G(X,X')= {1\over 4\pi}\left[  
{\cal G}_0(x,x') +2\sum_{n=1}^{\infty} \, \cos({n\over 2}(\tau -\tau'))\, {\cal
G}_n(x,x') \right]\, 
\ee
where ${\cal G}_n$ are ``radial'' Green functions which are solutions of
the following 1-dimensional problem
\be \n{n2.13}
\left[{d\over dx}(f{d\over dx}) -{\mu}^2 -{n^2\over 4f}-\xi \tilde{R}\right]{\cal
G}_n(x,x') = -\delta(x-x')\, .
\ee
By multiplying this equation by $f(x)$ and introducing new ''tortoise''
coordinate
\be\n{n2.14}
x_{*}\equiv \int_1^x{dx\over f(x)}=\ln(e^x-1)\, ,
\ee
one  can rewrite (\ref{n2.14}) as
\be\n{n2.15}
\left[{d^2\over dx^2_{*}}-U_n\right]{\cal G}_n(x,x') = -\delta(x_*-x'_*)\, ,
\ee
where
\be\n{n2.16}
U_n= {n^2\over 4} + U\, ,
\ee
and  is $U$ given by (\ref{2.11}).
The function $U_n$ has the asymptotic values $n^2/4$ and $\mu^2+n^2/4$ at
$x_*=-\infty$ and $x_*=\infty$, respectively. We denote by $R_n^>$
a solution of the homogeneous equation which has a decreasing
asymptotic $\exp(-\sqrt{\mu^2+n^2/4}\, x_*)$ at $x_*=\infty$, and  we denote
by $R_n^<$ a solution which has a non-increasing asymptotic
$\exp(n x_*/2)$ at $x_*=-\infty$. 

To obtain a solution $R_n$ of the homogeneous equation  we denote
\be\n{n2.17}
R_n(x)= z^{n/2}\, (1-z)^{\mu_n}\,  u(z)\, ,
\ee
where
\be\n{n2.18}
z= 1- e^{-x}\, ,\hspace{0.5cm}\mu_n=\sqrt{\mu^2+{n^2\over 4}}\, .
\ee
The function $u(z)$ obeys the hypergeometric equation (\ref{2.31}) with
\be\n{n2.19}
a=\mu_n +{n+1\over 2} -\beta~,\hspace{0.5cm}
b=\mu_n +{n+1\over 2} +\beta~,\hspace{0.5cm}
c=n +1~,\hspace{0.5cm}
\beta =\sqrt{{1\over 4}-\xi}\, .
\ee
Two linear independent solutions of this equations are
\be\n{n2.20}
R_n^<(x)= z^{n/2}\, (1-z)^{\mu_n} F(a,b; n+1;z) 
\, ,
\ee
and
\be\n{n2.20}
R_n^>(x)= z^{n/2}\, (1-z)^{\mu_n} F(a,b; \tilde{c};1-z)~, \hskip 1cm
\tilde{c}=2\mu_n+1 
\, .
\ee
The functions $R_n^<$ are regular at the horizon $x=0$ ($z=0$) and functions $R_n^>$ are decreasing at infinity $x=\infty$ ($z=1$).

To construct the Green function we need to calculate the Wronskian 
\be\n{n2.26}
W[R_n^>,R_n^<]={dR_n^<(x)\over dx_*}R_n^>(x)-{dR_n^>(x)\over
dx_*}R_n^<(x)\, .
\ee
The value of the Wronskian for the equation (\ref{n2.15}) does not
depend on the point. For this reason it is sufficient to calculate its
value at any point. It is convenient to perform
calculations at $x=1$. For this purpose we first transform the solution
$R_n^>$ into the form which has the same argument as $R_n^<$. Notice
that $\tilde{c}=a+b-n$. Using relation (15.3.12) of \cite{AbSt} we get
\ba
&&F(a,b;a+b-n;1-z)={\Gamma(n)\Gamma(\tilde{c})\over
\Gamma(a)\Gamma(b)}\, z^{-n}\, \sum_{k=0}^{n-1}\, {(a-n)_k\,
(b-n)_k\over k! (1-n)_k} \, z^k   \nonumber \\
\n{n2.27}
&&-{(-1)^n \Gamma(\tilde{c})\over
\Gamma(a-n)\Gamma(b-n)}\, \sum_{k=0}^{\infty} {(a)_k\,
(b)_k\over k! (k+n)_k} \, z^k\, \left[
\ln z -\psi(k+1)-\psi(k+n+1)+\psi(a+k)+\psi(b+k)\right]\, .
\ea
Here 
\be\n{n2.28}
(p)_k={\Gamma(p+k)\over \Gamma(p)}\, ,
\ee
and $\psi(\zeta)$ is the digamma function 
$\psi(\zeta)=\Gamma'(\zeta)/\Gamma(\zeta)$.
The relation (\ref{n2.27}) is valid for $|\mbox{arg} (z)|<\pi$ and $|z|<1$.
In the vicinity of $z=0$ the solutions have the following form
\ba
R_n^<(x)&=& z^{n/2}\, F_n^<(z)\, ,\n{n2.29} \\
R_n^>(x)&=& z^{-n/2}\, F_n^>(z)\, ,\hskip 0.5cm \mbox{for $n\ge 1$}\, ,\nonumber \\
R_0^>(x)&=& \ln z\, F_0^>(z)\, ,\hskip 0.85cm \mbox{for $n=0$}\, ,\n{n2.30}
\ea
where
\be\n{n2.31}
F_n^<(0)=1\, ,\hspace{0.5cm}F_{n\ge 1}^>(0)= {\Gamma(n)\, \Gamma(\tilde{c})\over
\Gamma(a)\, \Gamma(b)}\, ,\hspace{0.5cm}
F_0^>(0)=-{\Gamma(\tilde{c})\over \Gamma(a)\, \Gamma(b)}\, .
\ee
Using these relations we get
\be\n{n2.32}
W[R_n^>,R_n^<]\equiv {\cal W}_n= {\Gamma(n+1)\,
\Gamma\left(2\mu_n+1\right)\over
\Gamma(a)\, \Gamma(b)}\, .
\ee

Combining the above results we obtain the following expression for the
`radial' Green function
\be\n{n2.33}
{\cal G}_n(x,x')={1\over {\cal W}_n}\, R_n^>(x^>)R_n^<(x^<)\, ,
\ee
where
\be\n{n2.34}
x^> =\max (x,x')\, ,\hspace{0.5cm}x^< =\min (x,x')\, .
\ee

\subsection{Massless field case}

In case when $\mu =0$  the Green function can be obtained in an explicit form. For massless fields the modes $R^{>\atop <}_n$ simplify a little
and can be expressed in terms of associated Legendre functions
$P^n_{-\beta}$
\begin{eqnarray}
R_n^<(x)&=& [z(1-z)]^{n/2}~F\left(n+{1\over2}+\beta,n+{1\over2}-\beta;n+1;z\right)\\
&=&(-1)^n{\Gamma(n+1)\Gamma\left(-n+{1\over2}-\beta\right)\over
\Gamma\left(n+{1\over2}-\beta\right)}~P^{~n}_{-{1\over 2}-\beta}(1-2 z)~, \\
R_n^>(x)&=& [z(1-z)]^{n/2}~F\left(n+{1\over2}+\beta,n+{1\over2}-\beta;n+1;1-z\right)\\
&=&(-1)^n{\Gamma(n+1)\Gamma\left(-n+{1\over2}-\beta\right)\over
\Gamma\left(n+{1\over2}-\beta\right)}~P^{~n}_{-{1\over 2}-\beta}(-1+2 z)~, 
\end{eqnarray}
where $~z=1-{\mathrm e}^{-x}=(1+ {\mathrm e}^{x_*})^{-1}~$.
The Wronskian reads
\begin{eqnarray}
{\cal W}_n&=&{\Gamma(n+1)^2\over\Gamma\left(n+{1\over2}+\beta\right)
\Gamma\left(n+{1\over2}-\beta\right)}~,
\end{eqnarray}
Hence, the  Green functions ${\cal G}_n$ become 
\begin{eqnarray}
{\cal G}_n^(x,x')={(-1)^n\pi\over \cos(\pi \beta)}~
{\Gamma\left(-n+{1\over2}-\beta\right)\over\Gamma\left(n+{1\over2}-\beta\right)}
~P^{~n}_{-{1\over 2}-\beta}(-1+2 z^<)~P^{~n}_{-{1\over 2}-\beta}(1-2 z^>)~,
\end{eqnarray}
Substitution of these formulas into the expression (\ref{n2.12}) leads
to the series 
\begin{eqnarray}
&&G(X,X') = {1\over 4\cos(\pi \beta)}~\Big[~
P_{-{1\over 2}-\beta}(-1+2 z^<)~P_{-{1\over 2}-\beta}(1-2 z^>)  \nonumber \\
&&+ 2\sum_{n=1}^\infty (-1)^n\cos\left({n(\tau-\tau')\over 2}\right)
{\Gamma\left(-n+{1\over2}-\beta\right)\over\Gamma\left(n+{1\over2}-\beta\right)}
~P^{~n}_{-{1\over 2}-\beta}(-1+2 z^<)~P^{~n}_{-{1\over 2}-\beta}(1-2 z^>)   
\Big] ~.
\end{eqnarray}
Fortunately this summation can be performed completely 
\cite{Gradshteyn_Ryzhik} and we obtain 
a very simple explicit answer for the Green function of massless
nonminimal fields

\begin{eqnarray}
G(X,X') &=& {1\over 4\cos(\pi
\beta)}~P_{-{1\over 2}-\beta}(-\lambda)~,\label{Green}
\end{eqnarray}
where
\begin{eqnarray}
\lambda&=&(1-2z)(1-2z')+4\sqrt{zz'(1-z)(1-z')}\cos\left({\tau-\tau'\over
2}\right)~.
\end{eqnarray}

\section{Calculation of $\la \varphi^2\ra^{\ind{ren}}$}
\setcounter{equation}0

\subsection{$\la \varphi^2\ra^{\ind{ren}}$ on the horizon}
 
Before studying $\la \varphi^2(x)\ra^{\ind{ren}}$ in the black hole
exterior we  consider a
special case when the point $X$ is located on the horizon, $X=X_0$. In
this case $\la \varphi^2\ra^{\ind{ren}}$ can be calculated exactly. It
occurs because the Euclidean horizon is a fixed point of the Killing
vector field. For this reason the Green function $G(X,X_0)$ does not
depend on $\tau-\tau'$ and only $n=0$ contributes to this quantity in
the series (\ref{n2.12}). The functions $R^<_n$ are normalized so that
$R^<_0(x=0)=1$ (see (\ref{n2.29}) and (\ref{n2.31})). Using (\ref{n2.27})
and keeping only divergent and finite at the horizon terms in $R^>_0$,
we get
\be\n{n3.1}
G(X,X_0) \approx -{1\over 4\pi} \left[ 
\ln (1-e^{-x}) +2\gamma +\psi\left(\mu+{1\over 2}+\beta\right)
+\psi\left(\mu+{1\over 2}-\beta\right) 
\right] \, ,
\ee
where $\gamma=-\psi(1)$ is Euler's constant.

To get the renormalized value of $\la \varphi^2\ra$ we need first to
subtract the divergence from (\ref{n3.1}) and then to take the
coincidence limit $X\to X_0$
\be\n{n3.2}
\la \varphi^2(x=0)\ra^{\ind{ren}}=\lim_{X\to X_0}\,
(G(X,X_0)-G^{\ind{div}}(X,X_0))\, .
\ee
The divergent part of the Green function is (see Appendix)
\be\n{n3.3}
G^{\ind{div}}(X,X')=-{1\over 4\pi}\, \left[\ln\left({1\over
2}\mu^2\sigma(X,X')\right) +2\gamma\right]\, .
\ee
For $X'=X_0$, $\sigma(X,X_0)=l^2/2$ where the proper distance from the
horizon $l$ is
\be\n{n3.4}
l=x+2\ln \left(1+\sqrt{1-\exp(-x)}\right)\, .
\ee
After expansion of the right-hand sides of Eqs.(\ref{n3.1}-\ref{n3.3}) 
in powers of $x$ and cancellation of the divergences one gets
\be\n{n3.5}
\la \varphi^2(x=0)\ra^{\ind{ren}}= {1\over 4\pi}\left[ 2\ln \mu
-\psi\left(\mu+{1\over 2}+\sqrt{{1\over 4}-\xi}\right)-\psi\left(\mu+{1\over
2}-\sqrt{{1\over 4}-\xi}\right)\right]\, .
\ee
This answer is in agreement with the result of the paper 
\cite{FrMaSch-92},
where  $\la \varphi^2(x=0)\ra^{\ind{ren}}$ has been calculated for a
particular value of $\xi=1/4$.

\subsection{$\la \varphi^2\ra^{\ind{ren}}$ outside the horizon}

To obtain $\la \varphi^2\ra^{\ind{ren}}$ outside the horizon we use the
series representation for the Green function (\ref{n2.12}). Using mode
decomposition of the divergent part of the Green function $G^{\ind{div}}$
derived in the Appendix we get
\be\n{n3.6}
\la \varphi^2(x)\ra^{\ind{ren}}= {1\over 4\pi}\left[  
{\cal G}_0(x,x)+\ln(\mu^2 f)+2\gamma +
2\sum_{n=1}^{\infty} \, \left({\cal G}_n(x,x)-{1\over n}
\right)\right]\, , 
\ee
where ${\cal G}_n$ are given by (\ref{n2.33}).

For given parameters $\mu$ and $\xi$, plots of $\la
\varphi^2(x)\ra^{\ind{ren}}$ can be constructed by using numerical
calculations. Before discussing a behavior of $\la
\varphi^2(x)\ra^{\ind{ren}}$ we first check that the series in
(\ref{n3.6}) converges. To study large $n$ asymptotic of the terms of
the series we  use WKB approximation. Notice that solutions $R_n^{<}$
and $R_n^{>}$ can be written in the following form
\be\n{n3.7}
R_n(x)={1\over \sqrt{W_n(x)}}\, e^{\pm \int_{x_1}^{x}\, W_n(x')\, dx'} \, ,
\ee
where $W(x)$ is a solution of the equation
\be\n{n3.8}
W^2_n=U_n-{1\over 2}{W_n''\over W_n}+{3\over 4}{(W_n')^2\over W^2_n}\, ,
\ee
and $( )'=d( )/dx_*$.
In relation (\ref{n3.7}) signs $+$ and $-$ stand for solutions, $R_n^<$
and $R_n^>$, respectively. The Wronskian for these solutions,
(\ref{n2.26}), is 1, and hence partial radial Green functions in the
coincidence limit can be presented in the form
\be\n{n3.9}
{\cal G}_n(x,x)={1\over 2W_n(x)}\, .
\ee  
It should be emphasized that this result is exact provided $W_n(x)$ is
an exact solution of the equation (\ref{n3.8}). To obtain the large $n$
asymptotic of ${\cal G}_n(x,x)$ it is sufficient to use an approximate
solution 
\be\n{n3.10}
(W^{(1)}_n)^2=U_n={n^2\over 4}+f(x)\, (\mu^2+\xi\tilde{R})\, .
\ee
Iterations show that the omitted terms in equation (\ref{n3.8}) are of
higher order in $1/n$ expansion. In this approximation we get
\be\n{n3.11}
{\cal G}^{(1)}_n(x,x)={1\over n}-{2\over n^3}\, f(x)\, (\mu^2+\xi\tilde{R})\, .
\ee
This result demonstrates that the subtraction of $1/n$ term makes the
series in (\ref{n3.6}) convergent. Moreover, one can use the WKB result
to improve the convergence. By adding and subtracting ${\cal
G}^{(1)}_n(x,x)$ and using the definition of the Riemann zeta function
\be\n{n3.12}
\zeta(s)=\sum_1^{\infty} n^{-s}\, ,
\ee
one can present $\la\varphi^2(x)\ra^{\ind{ren}}$ in the form
\be\n{n3.13}
\la \varphi^2(x)\ra^{\ind{ren}}= {1\over 4\pi}\left[  
{\cal G}_0(x,x)+\ln(\mu^2 f)+2\gamma -4\zeta(3) f(x)\,
(\mu^2+\xi\tilde{R})+ \Delta {\cal G}(x,x)
\right]\, , 
\ee
\be\n{n3.14}
\Delta {\cal G}(x,x)=
2\sum_{n=1}^{\infty} \, \left({\cal G}_n(x,x)-{\cal G}^{(1)}_n(x,x)\right)\, .
\ee
The terms of the series decrease as $n^{-5}$. The convergence can still
be improved by using higher in $n^{-1}$ corrections in the WKB expansion
for ${\cal G}_n(x,x)$.

\subsection{Massless case}

Because in the massless case we know the Green function explicitly,
calculation of the $\la \varphi^2\ra^{\ind{ren}}$ is greatly simplified.
It is convenient to rewrite the answer Eq.(\ref{Green}) in the form
\begin{eqnarray}
G(X,X') &=& {1\over 4\cos(\pi
\beta)}~F\left({1\over 2}+\beta,{1\over 2}-\beta;~1;~{1+\lambda\over 2}\right)
\end{eqnarray}
For small point splitting $\lambda=1-2\epsilon$, where
$\epsilon\rightarrow 0$. Let us consider the separation in
z-direction $(~\tau-\tau'=0~)$ . Then the proper distance 
between the points reads
\begin{eqnarray}
\rho-\rho'=l(z,z')&=&2\ln(1+\sqrt{z})-\ln(1-z)-2\ln(1+\sqrt{z'})+\ln(1-z')\nonumber
\\ &=&{1\over (1-z)\sqrt{z}}~|z-z'|+O\left((z-z')^2\right)
\end{eqnarray}
and
\begin{eqnarray}
\epsilon&=&z+z'-2zz'-2\sqrt{zz'(1-z)(1-z')}\nonumber \\
&=&{1\over 4z(1-z)}(z-z')^2 +O\left((z-z')^3\right)={1-z\over4}~l^2 +
O(l^3)~.
\end{eqnarray}
In this limit the degenerate hypergeometric function is
\begin{eqnarray}
F\left({1\over 2}+\beta,{1\over 2}-\beta;1;1-\epsilon\right)
={\cos(\pi\beta)\over\pi}\left[-\ln(\epsilon)-2\gamma
-\psi\left({1\over 2}+\beta\right)-\psi\left({1\over 2}-\beta\right)\right]
+O(\epsilon^2)~.
\end{eqnarray}
Thus
\begin{eqnarray}
G(X,X')=-{1\over 4\pi}~\left[\ln(1-z)+\ln\left({l^2\over 4}\right)
+2\gamma+\psi\left({1\over 2}+\beta\right)+\psi\left({1\over 2}-\beta\right)\right]
+O(l^3)~.
\end{eqnarray}
Taking into account Eq.(\ref{n3.3})
\begin{eqnarray}
G(X,X')^{\mathrm {div}}&=&-{1\over 4\pi}~\left[\ln\left({\mu^2 l^2\over 4}\right)
+2\gamma\right]
\end{eqnarray}
we obtain
\begin{eqnarray}
\la \varphi^2(X)\ra^{\ind{ren}}&=&\lim_{X'\to X}\,
(G(X,X')-G^{\ind{div}}(X,X'))\\
&=&{1\over 4\pi}\left[-\ln(1-z)+\ln(\mu^2)
-\psi\left({1\over 2}+\beta\right)-\psi\left({1\over 2}-\beta\right)\right]\\
&=&{1\over 4\pi}\left[x+\ln(\mu^2)
-\psi\left({1\over 2}+\beta\right)-\psi\left({1\over 2}-\beta\right)\right]\, .
\end{eqnarray}
On the horizon (at $x=0$) this result, evidently, reproduces 
the massless limit $\mu\to 0$ of the Eq.(\ref{n3.5}), the mass $\mu$
playing the role of the infrared cut-off parameter.
\section{Conclusion}

We studied quantum effects for quantum nonminimal scalar field in a
two-dimensional black hole spacetime. We demonstrated that for a
string motivated black hole metric \cite{ManSenWad,Witten} the field
equation allows exact analytical solutions in terms of hypergeometric
functions. Using these solutions we obtained an explicit expression
for greybody factors and calculated Hawking radiation. We also
demonstrated that for negative values of nonminimal coupling
constant  $\xi$ the field besides usual scattering modes can have
bound states. The bound states for the scalar field of mass $m$ near
the black hole with the gravitational radius $r_0$ are present when
$-\xi\le\mu(\mu+1)$, where $\mu=mr_0$. These bound states lead to
instability. As a result of this the negative energy density in the
region in the black hole exterior grows exponentially. This effect
is  accompanied by exponentially growing positive energy fluxes
through the black hole horizon and to infinity. It should be
emphasized that this kind of instability occurs for any theory with
negative $\xi$ for solutions describing evaporating black holes, since
the gravitational radius $r_0\to 0$ and there is a moment of time when
the parameter $\mu$ meets the condition of formation of a bound state.
This result may indicate that 2D theories of the nonminimal scalar
field is inconsistent for negative $\xi$. 

The obtained explicit expressions for Green functions and
$\la\varphi^2\ra^{\ind{ren}}$ may be used to test an accuracy of
different analytical approximations developed for study of vacuum
polarization effects in the spacetime with the black hole.

\bigskip

\vspace{12pt}
{\bf Acknowledgments}:\ \  This work was  partly supported  by  the
Natural Sciences and Engineering Research Council of Canada. The
authors are grateful to the Killam Trust for its financial
support.

\appendix

\section{Divergencies}
\setcounter{equation}0

The divergent part of the Green function in two 
dimensions can be obtained by integrating 
Schwinger-DeWitt expansion of the heat kernel, 
\be\n{A.1}
K^{\mathrm {div}}(X,X'|s)
  =  \frac{D^{\frac{1}{2}}(X,X')}{4\pi s} 
     \exp{ \left\{ - \mu^2s -\frac{\sigma(X,X')}{2s} \right\} }
     \left[ 1 + \ldots \right]  \, ,
\ee
Here $\sigma$ is the half of the square of geodesic distance and $D$ is Van Fleak-Morette determinant. 
One has
\ba
G^{\mathrm {div}}(X,X')&=&
\frac{D^{\frac{1}{2}}(X,X')}{4\pi} \int_0^\infty ds~{1\over s} 
     \exp{ \left\{ - \mu^2s -\frac{\sigma(X,X')}{2s} \right\} }
\nonumber \\
&=& \frac{D^{\frac{1}{2}}(X,X')}{2\pi}
K_0(\mu\sqrt{2\sigma(X,X')})       |
\nonumber \\
&=& \frac{D^{\frac{1}{2}}(X,X')}{4\pi}\left[
  -\ln\left({\mu^2\sigma(X,X')\over 2}\right)-2\gamma 
  \right]\, . \nonumber
\ea
For splitting points in $\tau$ direction in the metric (\ref{2.1}) 
one has ($\Delta\tau =\tau-\tau'$)
\ba
2\sigma(X,X')&=&f(x) \Delta\tau^2 + O(\Delta\tau^4) = 
  8 f(x) \left[1-\cos\left({\Delta\tau\over 2}\right)\right] 
  + O(\Delta \tau^4) \, , \nonumber \\
D^{\frac{1}{2}}(X,X')&=&1 + O(\Delta\tau^2) \, .\nonumber
\ea
Thus
\be
G^{\mathrm {div}}(\tau,x;\tau',x)=-\frac{1}{4\pi}\left[
  \ln\left(2\left[1-\cos\left({\tau-\tau'\over 2}\right)\right]\right)
  +\ln\left({\mu^2 f}\right)+2\gamma 
  \right]\, .   
\ee
Using the relation
\begin{eqnarray*}
  \ln\left(2\left[1-\cos\left({\tau-\tau'\over
  2}\right)\right]\right)=\sum_{k=1}^\infty \left[-{2\over
  k}\cos\left(k{\tau-\tau'\over 2}\right)\right]   
\end{eqnarray*}
we get
\begin{eqnarray*}
  G^{\mathrm {div}}(\tau,x;\tau',x)&=&\frac{1}{4\pi}\left[
  -\ln\left({\mu^2 f}\right)-2\gamma
  +2\sum_{n=1}^\infty \cos\left({n\over 2}(\tau-\tau')\right)\times{1\over
  n}
  \right]   
\end{eqnarray*}
From this one can easily obtain for the Fourier components of the
UV divergent part of Green function
\begin{eqnarray*}
  {\cal G}^{\mathrm {div}}_0(x,x)&=&
  -\ln\left({\mu^2 f}\right)-2\gamma \, ,     \\ 
  {\cal G}^{\mathrm {div}}_n(x,x)&=&{1\over n}\, .
\end{eqnarray*}

\end{document}